\documentclass[14]{article}
\usepackage[T1]{fontenc}
\usepackage[ansinew]{inputenc}
\usepackage{array}
\usepackage{color}
\usepackage{amsmath}
\usepackage{amsxtra}
\usepackage{amstext}
\usepackage{amssymb}
\usepackage{latexsym}
\usepackage{dsfont}
\usepackage{graphicx,amssymb,amsfonts,latexsym,amsmath,amsthm,times,dsfont}
\usepackage{wrapfig}
\usepackage{footnote}
\usepackage{epsfig}
\begin{document}
\begin{center}
{\large\bf Stability of Logarithmic Bose-Einstein Condensate in Harmonic Trap}\\

\vspace{0.5cm}
{B. Bouharia}

\vspace{0.5cm}
{\it
Faculty of Sciences, University Mohammed First, BP 524,60000\\
Oujda, Morocco.
 }

\textrm{bouhariabrahim@yahoo.fr}\\
\end{center}

\hrulefill

{\bf Abstract}

\medskip
In this paper we investigate the stability of a recently introduced Bose-Einstein condensate (BEC) which involves logarithmic interaction between atoms. The Gaussian variational approach is employed to derive equations of motion for condensate widths in the presence of a harmonic trap. Then we derive the analytical solutions for these equations and find them to be in good agrement with numerical data. By analyzing deeply the frequencies of collective oscillations, and the mean-square radius, we find that the system is always stable for both negative and positive week logarithmic coupling. However, for strong interaction the situation is quite different: our condensate collapses for positive coupling and oscillates with fixed frequency for negative one. These special results remain the most characteristic features of the logarithmic BEC compared to that involving two-body and three-body interactions.

\vspace{1cm}
\section{Introduction}

Since its experimental observation in trapped alkali atomic vapors \cite{r1}, Bose-Einstein condensation (BEC) has largely attracted much attention from both theoretical and experimental viewpoints\cite{r2,r3,r4,r5}. Among all the recent developments, the study of collective excitations occupies an important position in
studying the properties of trapped  BEC. Due to the diluteness of  BEC, most theoretical works on collective excitations are mainly focused on considering the two-body interaction by means of the Gross-Pitaevskii equation (GPE) \cite{r6,r7,r8,r9}. A long side with the experimental progress with BECs in atomic waveguides and on the surface of atomic chips, which involve a strong compression of the traps, a significant increase of the density of BECs can be achieved; thus, three-body interaction may also play an important role in the process of collective excitations \cite{r10,r11,r12,r13}, and stability of trapped BEC \cite{r14,r15,r16}. Specially, it is reported in Ref.~\cite{r14} that the addition of three-body interaction can extend considerably the region of stability for a condensate, and allow the number of condensed atoms to increase even when the strength of the three-body force is very small. However, this correction remains insufficient given the fact that in higher densities regime nothing prevents the emergence of multi-body interactions. In this case, one should consider all the powers of $|\psi|^{2}$, $\psi$ being the condensate wave-function, which implies the usage of the non-polynomial functions that will necessarily appear in the wave equations for condensate.

Recently, K.G. Zloshchastiev \cite{r17} introduced a new quantum Bose liquid as a candidate structure of physical vacuum, this structure is described by a non-linear Schrodinger equation of a non-polynomial kind, namely the logarithmic one \cite{r18,r19,r20}. Taking into account small vacuum fluctuations, the author demonstrated by a simple model that the generated masses of the otherwise massless particles can be naturally expressed in terms of both elementary electrical charge and extensive length parameter of the non-linearity. Later in Ref.~\cite{r21}, it has been proved that the condensation governed by the logarithmic Schrodinger equation (LogSE) has several differences from that described by the GPE, particularly, it possesses the self-sustainability property: the logarithmic BEC tends always to form a Gaussian-type droplet- even in the absence of an external trapping potential. It has also been proved that the nature of elementary excitations depend strongly on the background density which changes the topological structure of their momentum space. In this paper, we attempt to explore more the properties of the logarithmic BEC, namely those related to collective excitations in a harmonic trap.
Our work is facilitated by variational approaches. Using the solutions in the Gaussian form, the LogSE is transformed into a set of differential equations about some parameters that characterize the condensate wave-function. It is confirmed that the analytical solutions of condensate widths equations conform to numerical data. By performing a stability analysis, based on two requirements for instability which gives us the critical points for collapse and which states that the BEC becomes unstable if the frequency of collective oscillations is zero or alternatively the mean-square radius of the condensate wave-function tends to zero in finite time, it is shown that our condensate is always stable, and that the collapse may occur only for large positive logarithmic interaction parameter. Finally, a comparison between the behavior of the logarithmic BEC and the classical one with two-body and three-body interactions indicates that the former is more stable than the latter.

The paper is organized as follows. In Sec.II, we present an analysis of
the spherically symmetric case, derive
the governing equation for the condensate width, and  discuss
stability criteria. In Sec.III, the same procedure is applied for the axially symmetric case, which reveals the effect of anisotropy. In Sec.IV, we report the numerical results of the present investigation. In Sec.V, we highlight the system of two-body and three-body interactions. Final section is our conclusion.

\section{Analysis of the Spherically Symmetric Case }

\subsection{Variational Approach and Governing Equations}

To begin, let us consider a boson gas with a fixed mean number of
particles $N$, moving in an isotropic harmonic
trap. If the particle density is considerably high and the temperature of the condensate
is low enough, the dynamics of the Bose-Einstein
condensed atoms can be described by the logarithmic Schrodinger equation (LogSE)\cite{r17,r18,r19,r20,r21}:
\begin{equation}\label{e1}
i\hbar \frac{\partial}{\partial t}\psi
=\left[-\frac{\hbar^{2}}{2m}\nabla^{2}+V-b
 \ln(a^{3} |\psi|^{2})\right]\psi\,.
\end{equation}

\noindent
where $\psi=\psi(r, t)$ being the condensate wave-function which is normalized to $N$, $m$ is the atom's mass. The parameter $b$ measures the strength of the non-linear interaction and
$a$ is needed to make the argument of the logarithm dimensionless.

Given a non-linear Schrodinger equation like (\ref{e1}) it is quite natural to tray a power expansion of the non-linear part $\ln(a^{3} |\psi|^{2})$, one will find the following
\begin{equation}\label{e2}
\ln(a^{3} |\psi|^{2})\approx-\frac{3}{2}+2 a^3 \left| \psi \right| ^2-\frac{a^6}{2}  \left| \psi \right| ^4\,.
\end{equation}

\noindent
The lowest order in $a^3 \left| \psi \right|^2$ gives the familiar Gross-Pitaevskii equation which describes only two-body interactions. By including the next term of the power one gets the so-called Gross-Pitaevskii-Ginzburg equation where also three-body interactions have to be taken into account (see for instance Ref.~\cite{r22}).
It is clear that the above expansion makes sense only when the density
is near the special value $1/a^{3}$ but in general
it is not always the case, for this reason the logarithmic non-linearity arises as an alternative
for describing the multi-body interactions.

Now $V\equiv V (r)$ is a static harmonic oscillator with spherical symmetry
given by $ V = \frac{1}{2}m \omega^{2} r^{2}$, $\omega$ is the radial
frequency of the isotropic trap.
By using dimensionless variables, $r \rightarrow l r$, $t\rightarrow t/\omega$, $l=
\sqrt{\hbar/m \omega}$, we redefine the condensate wave-function as
\begin{equation}\label{e3}\varphi(r,t)= a^{3/2}\psi(r,t)\,,\end{equation}

\noindent
such that
\begin{equation}\label{e4}4\pi\int_0^{\infty } r^2 |\varphi(r,t)|^{2} \, dr=N a^{3}=n\,,\end{equation}

\noindent
where $n$ is the reduced number of particles for the system.
Thus, the dimensionless equation corresponding to Eq.~(\ref{e1})
can be rewritten as
\begin{equation}\label{e5}
i\frac{\partial}{\partial t}\varphi
=\left[-\frac{1}{2}\nabla^{2}+\frac{1}{2}r^{2}-\lambda
 \ln(|\varphi|^{2})\right]\varphi\,,
\end{equation}

\noindent
where $\lambda$ is the dimensionless logarithmic parameter which describes the proportional relationship between the logarithmic coupling and the level spacing of the harmonic oscillator
\begin{equation}\label{e6}
\lambda=\frac{b}{\hbar \omega}\,.
\end{equation}

The chemical potential $\mu$ is given by the eigenvalue solutions
of Eq.~(\ref{e5}), with $\varphi(r, t) = \exp[-i(\mu t/2)]\phi(r)$:
\begin{equation}\label{e7}
\mu\phi=\left[-\nabla^{2}+ r^{2}-2\lambda \ln(|\phi|^{2})\right]\phi\,.
\end{equation}

In order to analyze the dynamics of logarithmic condensate in an harmonic trap, it is
convenient to follow the variational approach\cite{r7,r23}. Furthermore, we shall seek the solutions of Eq.~(\ref{e5}) in the Gaussian form \cite{r24}
\begin{equation}\label{e8}
\varphi(r,t)=A (t) \exp[-\frac{r^2}{2\xi (t)^{2}}+i \beta(t)r^2 +i \alpha(t)]\,,
\end{equation}

\noindent
where $A(t)$ is the amplitude, $\xi(t)$ is the width, $\alpha(t)$ is the
linear phase of the condensate, and $\beta(t)$ is the chirp parameter. The reduced number of particles $n$ is given by
\begin{equation}\label{e9}n=\pi^{3/2} A^2 \xi^3=constant\,.\end{equation}

The Lagrangian density generating Eq.~(\ref{e5}) in spherical coordinates is
\begin{equation}\label{e10}\mathcal{L}(\varphi)=\frac{i}{2}(\varphi\frac{\partial\varphi^*}{\partial t}-\varphi ^*\frac{\partial\varphi}{\partial t})+\frac{1}{2}|\frac{\partial\varphi}{\partial r}|^{2}+\frac{r^{2}}{2}|\varphi|^{2}-\lambda |\varphi|^{2}(\ln(|\varphi|^{2})-1)\,,\end{equation}

\noindent
By using Eqs.~(\ref{e8}) and (\ref{e10}), the effective Lagrangian density can be obtained,
\begin{equation}\begin{split}L_{eff}&=4\pi\int_0^{\infty } r^2\mathcal{L}(\varphi)dr\\
&=\pi ^{3/2}A^2\xi^{3}\Big[\frac{3}{4\xi^2}+\frac{\lambda}{2}\left(5-4 \ln(A)\right)\\& +\dot{\alpha}+\frac{3\xi^2}{4}\left(2\dot{\beta}+4\beta^2+1\right)\Big]\,.\label{e11}
\end{split}\end{equation}

\noindent
where the overhead dot represents time derivative.

Analyzing the corresponding Euler-Lagrange equations,
\begin{equation}\label{e12}\frac{\partial L_{eff}}{\partial q}-\frac{d}{d t}\frac{\partial L_{eff}}{\partial\dot{q}}=0\,,\end{equation}

\noindent
where $q$ stands for $A(t)$, $\xi(t)$, $\alpha(t)$ and $\beta(t)$ respectively, we obtain
\begin{equation}\label{e13}\pi ^{3/2}a^2\xi^3=n\,,\end{equation}

                         \begin{equation}\label{e14}
                      \frac{\dot{\alpha }}{2}+\frac{5 \beta ^2 \xi ^2}{2}+\frac{5 \lambda }{4}+\frac{5\dot{\beta } \xi ^2}{4}
   +\frac{1}{8 \xi ^2}+\frac{5 \xi ^2}{8}=\lambda  \ln (A)\,,
                         \end{equation}
                         \begin{equation}\label{e15}
                          \frac{\dot{\alpha}}{2}+\frac{3 \beta ^2 \xi ^2}{2}+\frac{3 \dot{\beta}\xi ^2 }{4}+\frac{3 \lambda
   }{4}+\frac{3 \xi ^2}{8}+\frac{3}{8 \xi ^2}=\lambda  \ln (A)\,,
                         \end{equation}
  \begin{equation}\label{e16}\dot{\xi}=2\xi\beta\,\end{equation}

\noindent
Eliminating $\dot{\alpha }$ between Eqs.~(\ref{e14}) and (\ref{e15}) one obtains
  \begin{equation}\label{e17}
                       2\dot{\beta}\xi=\frac{1}{\xi^{3}}-\xi-\frac{2\lambda}{\xi}-4\beta^{2}
                       \xi\,,
                         \end{equation}

\noindent
By combining Eqs.~(\ref{e16}) and (\ref{e17}) we get the following second-order differential equation for the evolution of the width
       \begin{equation}\label{e18}
                       \ddot{\xi}=-\xi+\frac{1}{\xi^{3}}-\frac{2\lambda}{\xi}\,,
                         \end{equation}

\noindent
Eq.~(\ref{e18}) illustrates the motion of a particle along the positive $\xi$ direction
in the effective potential $V(\xi)$ and can also be rewritten as :
\begin{equation}\label{e19}
                       \ddot{\xi}=-\frac{d V(\xi)}{d \xi}\,,
                         \end{equation}

                       \noindent
where
\begin{equation}\label{e20}
                       V(\xi)=\frac{\xi^{2}}{2}+\frac{1}{2\xi^{2}}+2\lambda \ln(\xi)\,,
                         \end{equation}

\noindent
The interpretation of Eq.~(\ref{e18}) is straightforward. The first term of the right-hand side of this equation corresponds to the attractive effect of the harmonic potentiel, the second is proportional to $\xi^{-3}$ and is related to the dispersive effect caused by kinetic energy, and the third one is proportional to $\xi^{-1}$ and coming from the non-linear logarithmic interaction. In order to understand the condensate dynamics, we must focus on the study of Eq.~(\ref{e18}). At the first sight we can see that it is a non-linear differential equation. Therefore, though this equation tends to simplify the original problem of Eq.~(\ref{e1}), it is still complex and non-integrable considering that the only conserved quantity is the Hamiltonian. A formal solution of Eq.~(\ref{e18}) is obtained by taking the Hamiltonian $H$ of the point particle
\begin{equation}\label{e21}
                       H=\frac{1}{2}(\frac{d\xi}{dt})^{2}+V(\xi)\,,
                         \end{equation}

\noindent
after integrating we get
\begin{equation}\label{e22}
                       t=\int_{\xi_{0}}^{\xi}\frac{d \xi^{'}}{\sqrt{2\big[H-\frac{\xi^{'2}}{2}-\frac{1}{2\xi^{'2}}-2\lambda \ln(\xi^{'})\big]}}\,,
                         \end{equation}

\noindent
However, this solution doesn't provide sufficient information about the condensate dynamics. Hence we infer that
the present problem can be solved only by considering some approximations which will be discussed
below in more detail.
\begin{figure}[t]
\centerline{\psfig{file=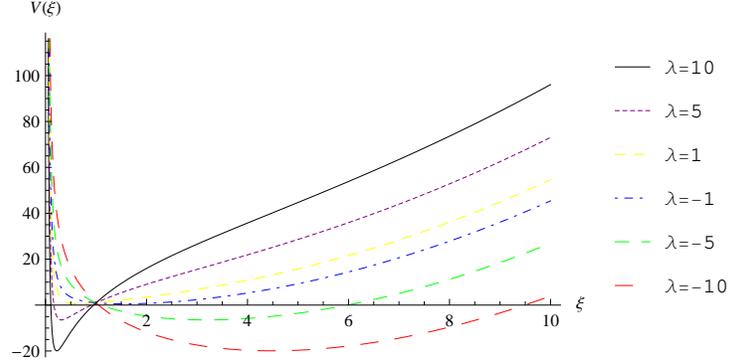}}
\vspace*{8pt}\caption{The effective potential $V(\xi)$ versus
the width $\xi$ for different values of logarithmic parameter $\lambda$}
\end{figure}

Now let us analyze the nature of the effective potential of Eq.~(\ref{e20}). Fig. 1
depicts the potential energy curves as a function of $\xi$ for different values of the dimensionless logarithmic parameter $\lambda$. We notice that the potential has a simple shape and contains for each case one local
minimum, which means that the only reasonable motion of the system is periodic
(anharmonic) oscillation around the minimum of the effective potential
$V(\xi)$.

\subsection{Oscillations and Stability of the Logarithmic BEC}

Obviously Eq.~(\ref{e18}) shows that the internal dynamic represented by the evolution of the width of the condensate is affected by the logarithmic interaction between atoms. The equilibrium width $\xi_{0}$ which corresponds to the stationary states of the condensates can be calculated by setting the gradient of the potential Eq.~(\ref{e20}) equal to zero
\begin{equation}\label{e23}
                       \frac{d V(\xi)}{d \xi}|_{\xi=\xi_{0}}=0\,,
                         \end{equation}
\noindent
which yields
 \begin{equation}\label{e24}
                       \xi_{0}=\frac{1}{\xi_{0}^{3}}-\frac{2\lambda}{\xi_{0}}\,,
                       \end{equation}

\noindent
Eq.~(\ref{e24}) can be rewritten in polynomial form as

\begin{equation}\label{e25}
                       \xi_{0}^{4}+2\lambda\xi_{0}^{2}-1=0\,,
                       \end{equation}

\noindent
Note that for given $\lambda$ only positive real solutions $\xi_{0}$ from Eq.~(\ref{e25})
represents physically realized equilibrium width, in our case the relevant solution is given by
\begin{equation}\label{e26}
\xi_{0}=\Big[\left(\lambda ^2+1\right)^{1/2}-\lambda\Big]^{1/2}\,.\end{equation}

\noindent
Expanding Eq.~(\ref{e18}) around the equilibrium width, we can obtain the dynamical equation of the width
 \begin{equation}\label{e27}
                       \xi=\xi_{0}+ A\sin(\omega_{r} t+\theta)\,.
                       \end{equation}

\noindent
where $A$, $\theta$ are real constant and $\omega_{r}$ is the frequency of the collective oscillations (in units of $\omega$) which is given by
\begin{equation}\label{e28}
                       \omega_{r}=\sqrt{\frac{d^{2}V(\xi)}{d^{2}\xi}\mid_{\xi=\xi_{0}}}=\Big(1+\frac{3}{\xi_{0}^{4}}-\frac{2\lambda}{\xi_{0}^{2}}\Big)^{1/2}\,,
                       \end{equation}

\noindent
From Eq.~(\ref{e26}) one gets
\begin{equation}\label{e29}
                       \omega_{r}=\Big\{1+\frac{3}{[\left(\lambda ^2+1\right)^{1/2}-\lambda]^{2}}-\frac{2\lambda}{\left(\lambda ^2+1\right)^{1/2}-\lambda}\Big\}^{1/2}\,.
                       \end{equation}

\noindent
Note that the solution Eq.~(\ref{e26}) is stable only if the frequencies of
collective modes are real, otherwise the solution is unstable. The frequency of collective oscillations, $\omega_{r}$, is among the relevant quantities usually used in literature to analyze the stability of the condensate. One of the requirements for instability of the BEC is that $\omega_{r}$ is zero \cite{r24}, this condition gives us the critical points at which the system collapses. From Eq.~(\ref{e29}) one can derive that these critical points are given by
\footnote{Note that for the BEC with two-body and three-body interaction the corresponding parameter of interaction can be complex. The imaginary part describes the effect of inelastic collisions on the dynamics of BEC \cite{r25,r26}. Nevertheless this interpretation doesn't correspond with our case as it leads to non-physical complex values of the width.}
\begin{equation}\label{e30}
\lambda_{c}=\pm i\,.
\end{equation}

\noindent
However, this complex values don't have any physical meaning in view of the fact that $\lambda$ is the dimensionless logarithmic parameter, which describes the proportional relationship between the logarithmic coupling and the level spacing of the harmonic oscillator that must take real values. Thus, our condensate remains always stable and the collapse will not take place in the case of small strength interaction. Fig.~2(a) shows the frequency of collective oscillations as functions of the logarithmic parameter. We notice that the frequency of collective oscillations presents a relative dependance on the logarithmic parameter $\lambda$. In fact, we see that this frequency decreases with the decreasing of $\lambda$ and tends almost to 1.4 for a large negative value of $\lambda$.
\begin{figure}[t]
\hspace{0.2cm}\begin{minipage}{.4\linewidth}
 \centerline{\psfig{file=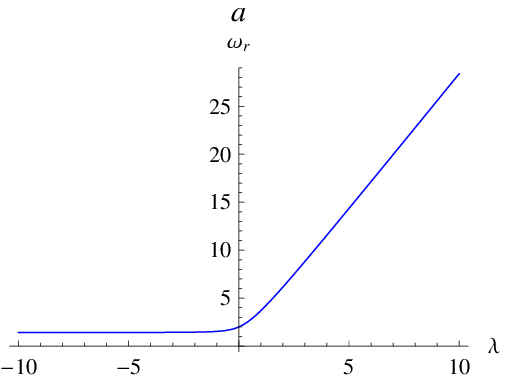}}
\end{minipage}
\hspace{1.5cm}
\begin{minipage}{.4\linewidth}
 \centerline{\psfig{file=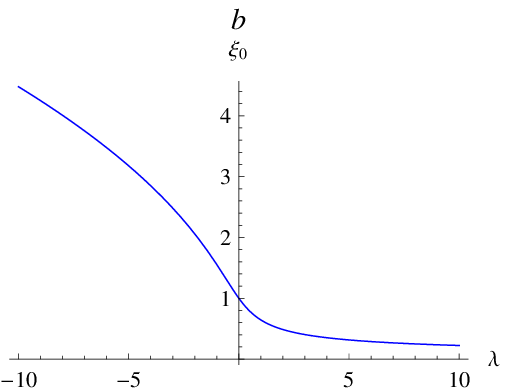}}
\end{minipage}
\vspace*{8pt}\caption{(a) The oscillation frequency $\omega_{r}$ and (b) The stable equilibrium width $\xi_{0}$ versus the logarithmic parameter $\lambda$}
\end{figure}

The mean-square radius is another issue of stability, briefly
speaking, collapse occurs as the mean-square radius of
the condensate wave-function, $\langle r^{2}\rangle=n^{-1}\int r^{2} |\varphi|^{2} dr$, tends
to zero in finite time, (see, e.g., \cite{r24,r27}), in the present units the corresponding expression is given by
\begin{equation}\label{e31}\langle r^{2}\rangle=\frac{3}{2}\xi^2\,.\end{equation}

\noindent
taking into account that $\phi$ governed by Eq.~(\ref{e7}) realizes an extremum of the Hamiltonian and using the following scaling \cite{r24,r27}
\begin{equation}\label{e32}\phi=\tilde{l}^{-3/2}\chi(\tilde{r})\,, \hspace{1cm}with\hspace{1cm}\tilde{r}\equiv\frac{r}{\tilde{l}}\,, \end{equation}

\noindent
where $\tilde{l}$ plays the role of a Lagrange multiplier, it is easy to express the Hamiltonian and the chemical potential in term of $\langle r^{2}\rangle$, for the Hamiltonian we obtain
\begin{equation}\label{e33}H(\tilde{l})=\frac{X_{s}}{2\tilde{l}^{2}}+\frac{\tilde{l}^{2}\langle \tilde{r}^{2}\rangle}{2}-\lambda (Y_{s}-Z_{s})\,,\end{equation}

\noindent
where
\begin{equation}\label{e34}X_{s}=-\int\chi^{\dag}\Delta\chi d^{3}\tilde{r}\,,\hspace{1cm}\langle \tilde{r}^{2}\rangle=\int \tilde{r}^{2}|\chi|^{2} d^{3}\tilde{r}\,,\end{equation}
\begin{equation}\label{e35}Y_{s}=\int|\chi|^{2}\ln(\tilde{l}^{-3}|\chi|^{2})d^{3}\tilde{r}\,,\hspace{1cm} Z_{s}=\int|\chi|^{2} d^{3}\tilde{r}\,,\end{equation}

\noindent
The minima of $H$ are given by the roots of the identity
\begin{equation}\label{e36}\frac{\delta H(\tilde{l})}{\delta \tilde{l}}\mid _{\tilde{l}=1}=0\,,\end{equation}

\noindent
then we obtain the following characteristic relation
\begin{equation}\label{e37}X_{s}-\langle r^{2}\rangle-3\lambda Z_{s}=0\,.\end{equation}

\noindent
and the Hamiltonian becomes
\begin{equation}\label{e38}H=\frac{5}{6}X_{s}+\frac{1}{6}\langle r^{2}\rangle-\lambda Y_{s}\,.\end{equation}

\noindent
In the same way, we can obtain the corresponding chemical potential as
\begin{equation}\label{e39}\mu=2\langle r^{2}\rangle-\lambda (2Y_{s}-3Z_{s})\,.\end{equation}

\noindent
from Eq.~(\ref{e31}) one gets
\begin{equation}\label{e40}\mu=3\xi^{2}-\lambda (2Y_{s}-3Z_{s})\,.\end{equation}

\noindent
From Eq.~(\ref{e31}) it is clear that an examination of the effect of the mean-square radius on the stability of the logarithmic BEC system, can be attained by considering the equilibrium width. Fig.~2(b) shows the equilibrium width as functions of the logarithmic parameter $\lambda$. We see that the equilibrium width decreases with increasing $\lambda$ and tends to zero for a large positive value, this means that it is only for a sufficiently large positive logarithmic interaction between atoms that collapse can take place.

\section{Axially Symmetric Case}

Analogous to the procedure of last section, we consider the logarithmic equation in dimensionless variables for an axially symmetric case. The governed equation for the logarithmic condensate wave function $\varphi(\rho, z,t)$ at radial position $\rho$, axial position z and time t can be written as
\begin{equation}\label{e41}
i\hbar \frac{\partial}{\partial t}\varphi
=\Big[-\frac{1}{2\rho}\frac{\partial}{\partial \rho}\Big(\rho\frac{\partial}{\partial \rho}\Big)-\frac{1}{2}\frac{\partial^{2}}{\partial z^{2}}+\frac{1}{2}\Big(\rho^{2}+\lambda_{1}^{2}z^{2}\Big)-\lambda_{2}
\ln\Big(|\varphi|^{2}\Big)\Big]\varphi\,,
\end{equation}

\noindent
where $V(\rho, z) = \rho^{2}+\lambda_{1}^{2}z^{2}$ is the axial trap. Here length and time are
expressed in units of $l(\equiv\sqrt{\frac{\hbar}{m \omega}})$ and $\omega^{-1}$ respectively, with $\omega$ the radial trap frequency, $\lambda_{1}=\omega_{z}/ \omega$ being constant describing the anisotropy of the trap which denotes the ratio of the frequency
along the $z$ direction $\omega_{z}$ to the radial frequency
$\omega$, and $\lambda_{2} =\frac{b}{\hbar \omega}$ is the dimensionless logarithmic parameter.

\noindent
We therefore try the product ansatz
\begin{equation}\label{e42}\varphi(\rho,z,t)=A (t)\prod_{\varrho=\rho,z}\exp[-\frac{\varrho^2}{2\xi_{\varrho}(t)^{2}}+i\beta_{\varrho}(t)\varrho^2 +\frac{i \alpha_{\varrho}(t)}{2}]\,,\end{equation}

\noindent
with normalization
\begin{equation}\label{e43}2\pi\int_{-\infty}^{\infty}\, dz\int_0^{\infty } \rho|\varphi(\rho,z,t)|^{2} \, d\rho=N a^{3}=n\,,\end{equation}

\noindent
The Lagrangian density in this case is
\begin{equation}\label{e44}\hspace{-0.1cm}\mathcal{L}(\varphi)=\frac{i}{2}(\varphi\frac{\partial\varphi^*}{\partial t}-\varphi ^*\frac{\partial\varphi}{\partial t})+\frac{1}{2}(|\frac{\partial\varphi}{\partial \rho}|^{2}+|\frac{\partial\varphi}{\partial z}|^{2})+\frac{1}{2}(\rho^{2}+\lambda_{1}^{2}z^{2})|\varphi|^{2}-\lambda_{2} |\varphi|^{2}(\ln(|\varphi|^{2})-1)\,,\end{equation}

\noindent
whereas the effective Lagrangian is given by
\begin{equation}\begin{split}L_{eff}&=2\pi\int_{-\infty}^{\infty}\, dz\int_0^{\infty } \rho\mathcal{L}(\varphi)d\rho\\
&=\pi ^{3/2}A(t)^2\xi_{\rho}^{2}\xi_{z}\Big[\frac{1}{2\xi_{\rho}^2}+\frac{1}{4\xi_{z}^2}+\frac{\lambda_{2}}{2}\left(5-4 \ln(A)\right)+\frac{1}{2}(\dot{\alpha_{\rho}}+\dot{\alpha_{z}})\\&+\frac{\xi_{\rho}^2}{2}\left(2\dot{\beta_{\rho}}+4\beta_{\rho}^2+1\right)
+\frac{\xi_{z}^2}{4}\left(2\dot{\beta_{z}}+4\beta_{z}^2+\lambda_{1}^{2}\right)\Big]\,.\label{e45}
\end{split}\end{equation}

\noindent
As in the isotropic case one can write the Euler-Lagrange equations
for $\alpha_{i}(t)$, $A(t)$, $\beta_{i}(t)$, and $\xi_{i}(t)$. After some algebra one can eliminate the variables $\alpha_{i}(t)$, $A(t)$, and $\beta_{i}(t)$ from these equations and obtain the following second-order differential
equations for the evolution of the condensate widths

\begin{figure}
\hspace{1.2cm}\begin{minipage}{.4\linewidth}
 \centerline{\psfig{file=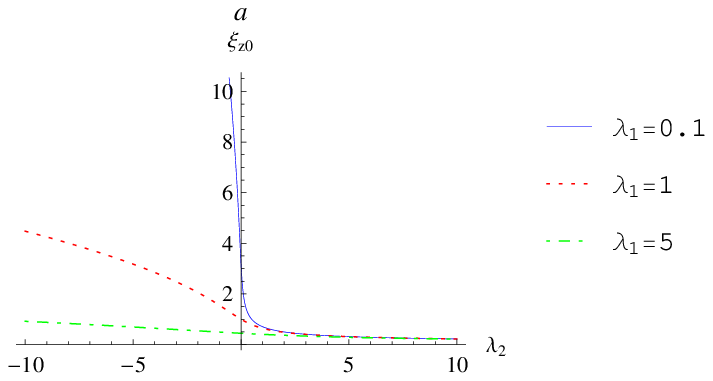}}
\end{minipage}
\hspace{0.5cm}
\begin{minipage}{.4\linewidth}
 \centerline{\psfig{file=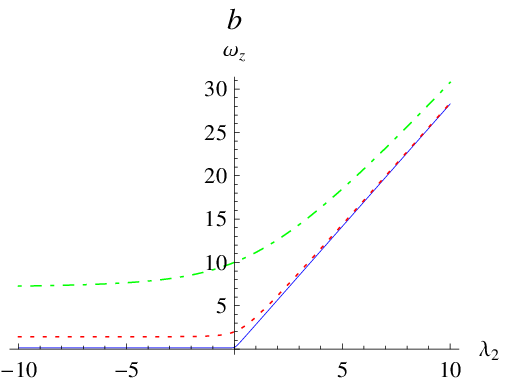}}
\end{minipage}
\vspace*{8pt}\caption{(a) The equilibrium width $\xi_{z0}$, (b) The frequency $\omega_{z}$, versus
the dimensionless logarithmic parameter $\lambda_{2}$ for several values of $\lambda_{1}$}
\end{figure}

\begin{equation}\label{e46}
       \ddot{\xi_{\rho}}=\frac{1}{\xi_{\rho}^{3}}-\xi_{\rho}-\frac{2\lambda_{2}}{\xi_{\rho}}\,,
                         \end{equation}
\begin{equation}\label{e47}
                       \ddot{\xi_{z}}=\frac{1}{\xi_{z}^{3}}-\lambda_{1}^{2}\xi_{z}-\frac{2\lambda_{2}}{\xi_{z}}\,.
                         \end{equation}

\noindent
Now the equilibrium widths equations are given by
\begin{equation}\label{e48}
       \xi_{\rho0}=\frac{1}{\xi_{\rho0}^{3}}-\frac{2\lambda_{2}}{\xi_{\rho0}}\,,
                         \end{equation}
\begin{equation}\label{e49}
                     \lambda_{1}^{2}\xi_{z0}=\frac{1}{\xi_{z0}^{3}}-\frac{2\lambda_{2}}{\xi_{z0}}\,.
                         \end{equation}
and their solutions are respectively
\begin{equation}\label{e50}
\xi_{\rho0}=\Big[\left(\lambda_{2}^2+1\right)^{1/2}-\lambda_{2}\Big]^{1/2}\,.\end{equation}

\begin{equation}\label{e51}
\xi_{z0}=\frac{[(\lambda_{1}^2+\lambda_{2}^2)^{1/2}-\lambda_{2}]^{1/2}}{\lambda{_1}}
\end{equation}
\noindent
Using Eqs.~(\ref{e50}) and (\ref{e51}), we find that the frequencies of oscillations around these equilibrium widths are given respectively by
\begin{equation}\label{e52}
                       \omega_{\rho}=\Big\{1+\frac{3}{[\left(\lambda_{2} ^2+1\right)^{1/2}-\lambda_{2}]^{2}}-\frac{2\lambda_{2}}{\left(\lambda_{2} ^2+1\right)^{1/2}-\lambda_{2}}\Big\}^{1/2}\,,
                       \end{equation}
\begin{equation}\label{e53}
                       \omega_{z}=\lambda_{1}\Big\{1+\frac{3\lambda_{1}^{2}}{[\left(\lambda_{1}^2+\lambda_{2} ^2\right)^{1/2}-\lambda_{2}]^{2}}-\frac{2\lambda_{2}}{\left(\lambda_{1}^2+\lambda_{2} ^2\right)^{1/2}-\lambda_{2}}\Big\}^{1/2}\,.
                       \end{equation}

\noindent
From Eq.~(\ref{e53}), we find that the values of $\lambda_{2}$ which lead to the axial collapse are given by
\begin{equation}\label{e54}
                       \lambda_{2}=\pm i\lambda_{1}\,.
                       \end{equation}
\noindent
Which are again non-physical complex values. Figs.~3(a) and 3(b) demonstrate respectively the equilibrium width and the frequency of axial mode as functions of the logarithmic parameter for different values of the anisotropy parameter. We see that the width decreases and the frequency increases with increasing either the positive logarithmic parameter or the anisotropy. In addition, if the interaction parameter is positive and very small, the anisotropy makes a great effect on both width and frequency, but when the interaction parameter becomes larger, this effect disappears and the width as well as the frequency are almost determined by $\lambda_{2}$. Figs.~3(b) shows also that if the negative value of $\lambda_{2}$ gets larger, the frequency reaches a relative constant value depending on the parameter of anisotropy $\lambda_{1}$. Furthermore, with much larger anisotropy, the equilibrium width tends to zero, which practically represents a collapsed condensate.

\section{Numerical Results}

The main results of our investigation are set out in graphic
forms. The graphs presented here show the evolution of the condensate width in spherically (Figs.~4(a)-4(f)) and axially (Figs.~5(a)-5(d)) symmetric traps when Eqs.~(\ref{e18}) and (\ref{e47}) are solved by using a numerical routine. This numerical routine was implemented through the use of Mathematica for some different values of the logarithmic parameter $\lambda$ (or $\lambda_{2}$) and the anisotropy parameter $\lambda_{1}$.
\begin{figure}
\begin{minipage}{.4\linewidth}
 \centerline{\psfig{file=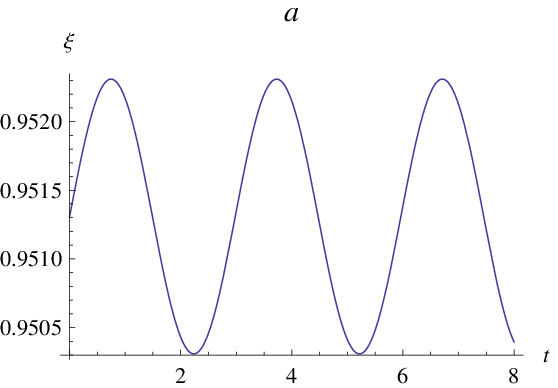,width=1.8in}}
\end{minipage}
\hspace{1cm}
\begin{minipage}{.4\linewidth}
 \centerline{\psfig{file=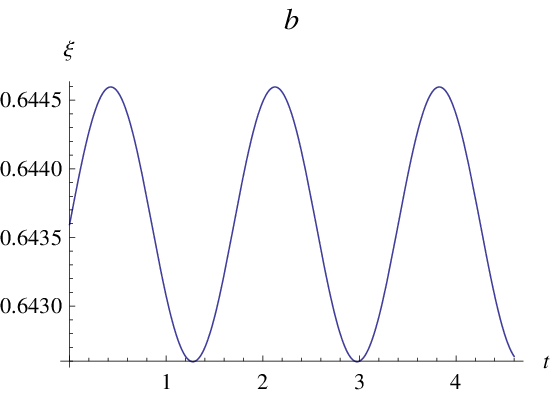,width=1.8in}}
\end{minipage}

\begin{minipage}{.4\linewidth}
 \centerline{\psfig{file=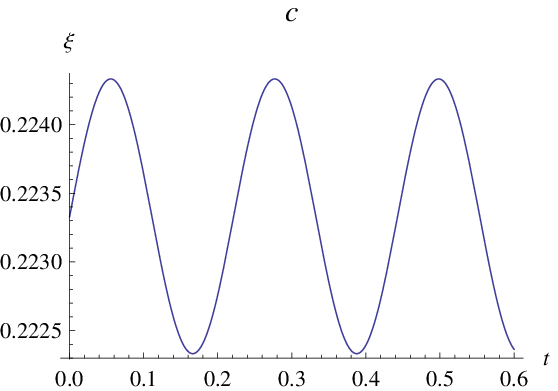,width=1.8in}}
\end{minipage}
\hspace{1cm}
\begin{minipage}{.4\linewidth}
 \centerline{\psfig{file=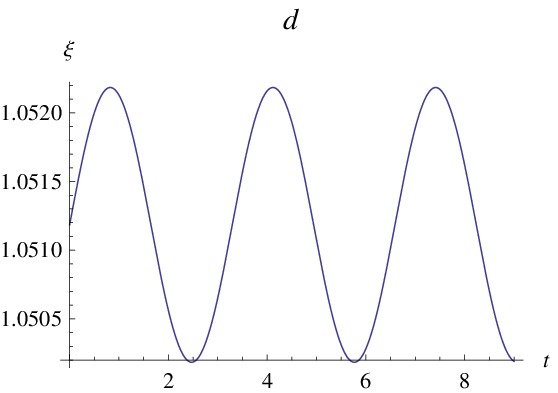,width=1.8in}}
\end{minipage}

\begin{minipage}{.4\linewidth}
 \centerline{\psfig{file=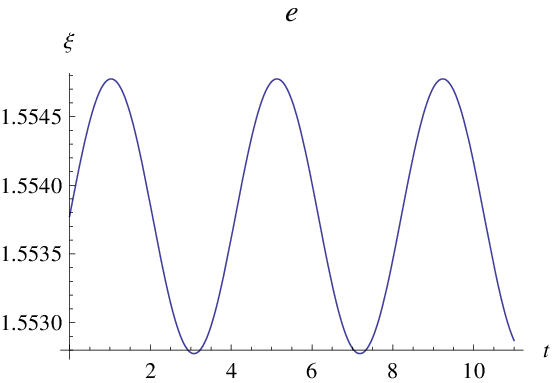,width=1.8in}}
\end{minipage}
\hspace{1cm}
\begin{minipage}{.4\linewidth}
 \centerline{\psfig{file=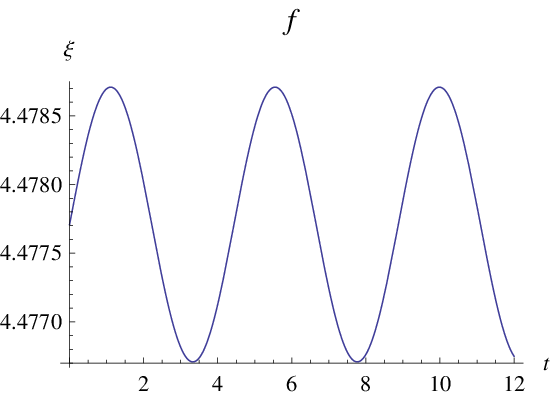,width=1.8in}}
\end{minipage}
\vspace*{8pt}\caption{Evolution of the condensate width in spherically symmetric case (a) $\xi_{0}= 0.951308$  and $\lambda=0.1$, (b) $\xi_{0}= 0.643$  and $\lambda=1$, (c) $\xi_{0}=0.223$  and $\lambda=10$, (d) $\xi_{0}=1.051$  and $\lambda=-0.1$, (e) $\xi_{0}=1.55377$  and $\lambda=-1$, and (f) $\xi_{0}= 4.477$  and $\lambda=-10$}
\end{figure}

It is clear that these numerical results are in a good agreement with the analytical ones. Indeed, for the spherically symmetric case (Figs.~4(a)-4(f)) the condensate width performs in all cases small oscillations about the equilibrium width, furthermore, by increasing $\lambda$ the period decreases (the frequency increases), whereas, the period is equal almost to 5.5 (the frequency $\simeq$ 1.14 ) if $\lambda =-10$. Another aspect of this agreement appears in the value of the frequency itself, for example the frequency drawn from Eq.~(\ref{e28}) for $\lambda=10$ is 28.39 and from Fig.~4(c) we can find that it is equal to 28.45.

\begin{figure}
\begin{minipage}{.4\linewidth}
 \centerline{\psfig{file=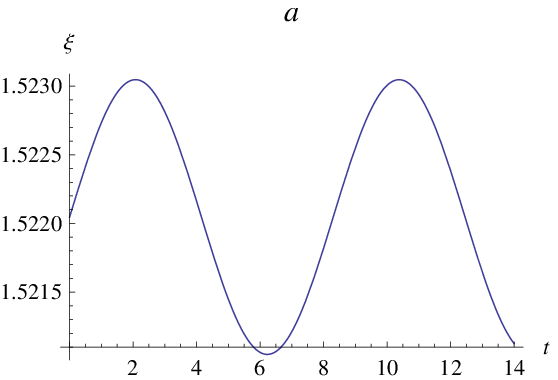,width=1.8in}}
\end{minipage}
\hspace{1cm}
\begin{minipage}{.4\linewidth}
 \centerline{\psfig{file=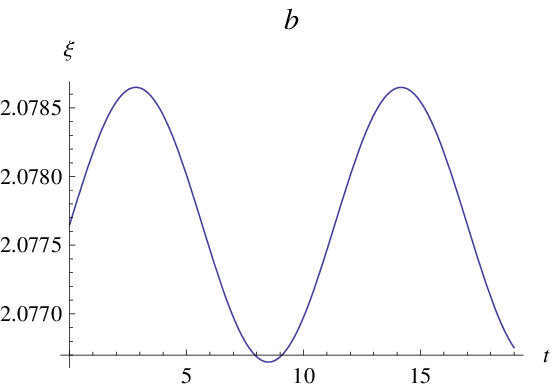,width=1.8in}}
\end{minipage}

\begin{minipage}{.4\linewidth}
 \centerline{\psfig{file=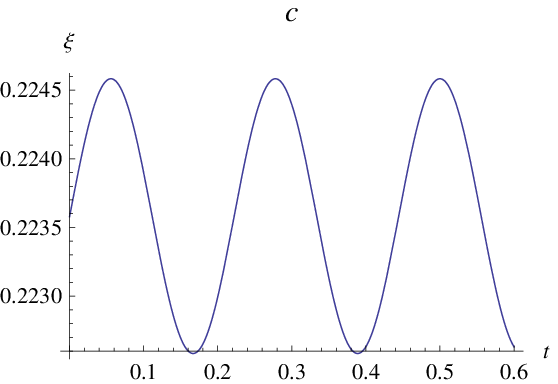,width=1.8in}}
\end{minipage}
\hspace{1cm}
\begin{minipage}{.4\linewidth}
 \centerline{\psfig{file=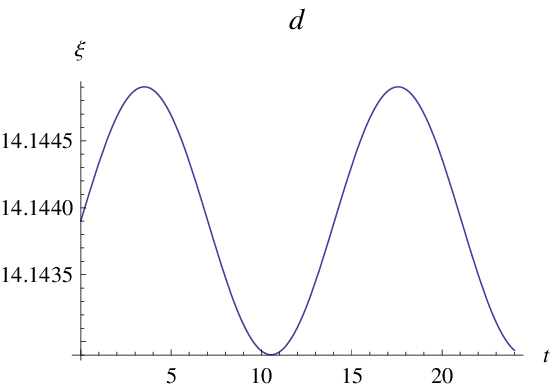,width=1.8in}}
\end{minipage}
\vspace*{8pt}\caption{Evolution of the condensate width in axially symmetric case (a) $\lambda_{1}= 0.1$  and  $\lambda_{2}=0.1$, (b) $\lambda_{1}= 0.1$  and  $\lambda_{2}=-0.1$, (c) $\lambda_{1}= 0.1$  and $\lambda_{2}=10$ (d) $\lambda_{1}= 0.1$  and $\lambda_{2}=-10$}
\end{figure}

The same agreement is also observed for the case of axially symmetric trap, namely if we compare on the one hand Fig.~4(a) with Fig.~5(a) and Fig.~4(d) with Fig.~5(b), and on the other hand Fig.~4(c) with Fig.~5(c) and Fig.~4(f) with Fig.~5(d), we find that the frequency increases for both positive and negative interaction parameter with the increasing of the anisotropy, but when the positive interaction parameter is much larger the anisotropy has no great effect on the frequency.

\section{Bose-Einstein Condensate with Two-body and Three-body Interactions}

The present section reviews mostly the results of earlier works concerning the BEC with two-body and three-body interactions \cite{r24,r28} and gives a comparison between this system and the logarithmic one. We should start writing down the governed equation in a dimensionless form
\begin{equation}\label{e55}
i\frac{\partial}{\partial t}\varphi
=\left[-\frac{1}{2}\nabla^{2}+V+\lambda_{2}
 |\varphi|^{2}+\lambda_{3}
 |\varphi|^{4}\right]\varphi\,,
\end{equation}

\noindent
In Eq.~(\ref{e55}) we are assuming dimensionless variables: the unit of length is
$l=\sqrt{\hbar/m\omega}$; and the unit of time is $1/\omega$. $V\equiv V(r, z)$ is a static trap
potential, that we assume a harmonic oscillator with
axially symmetry given by $V = \frac{1}{2}(r^{2}+\lambda_{1}^{2}z^{2})$, $\lambda_{1}=\omega_{z}/ \omega$ is the anisotropy ratio. $\lambda_{2}$ and $\lambda_{3}$ are respectively the parameters of the two-body and
three-body interactions which are considered reals. In particular $\lambda_{2}$ is proportional to the
two-body scattering length $a_{sc}$ and given by $\lambda_{2} = 4\pi a_{sc}/\hbar\omega l^{3}$.

Next we give the analogs of the above constructions. Using the same apparatus of the last section, we write down the equations for the widths

\begin{equation}\label{e56}
\ddot{\xi_{r}}=\frac{1}{\xi_{r}^{3}}-\xi_{r}+\frac{P}{\xi_{r}^{3}\xi_{z}}+\frac{K}{\xi_{r}^{5}\xi_{z}^{2}}\,,
                         \end{equation}
\begin{equation}\label{e57}
\ddot{\xi_{z}}=\frac{1}{\xi_{z}^{3}}-\lambda_{1}^{2}\xi_{z}+\frac{P}{\xi_{r}^{2}\xi_{z}^{2}}+\frac{K}{\xi_{r}^{4}\xi_{z}^{3}}\,,
                         \end{equation}
\noindent
Here, we have introduced the dimensionless two-body and three-body interaction strengths given respectively by

\begin{equation}\label{e58}
    P= \frac{\lambda_{2} N}{(2\pi)^{3/2}}\,,  \hspace{1.5cm} K= \frac{4 \lambda_{3}N^{2}}{9\sqrt{3}\pi^{3}}\,,
\end{equation}

\noindent
where $N$ is the number of atoms. The equilibrium widths are given by

\begin{equation}\label{e59}
\frac{1}{\xi_{r0}^{3}}-\xi_{r0}+\frac{P}{\xi_{r0}^{3}\xi_{z0}}+\frac{K}{\xi_{r0}^{5}\xi_{z0}^{2}}=0\,.
                         \end{equation}
\begin{equation}\label{e60}
\frac{1}{\xi_{z0}^{3}}-\lambda_{1}^{2}\xi_{z0}+\frac{P}{\xi_{r0}^{2}\xi_{z0}^{2}}+\frac{K}{\xi_{r0}^{4}\xi_{z0}^{3}}=0\,.
                         \end{equation}

\noindent
Now let us analyze the stability of the present system with the emphasis on the simple case of isotropic trap. In this case we have similarity between Eqs.~(\ref{e56}) and (\ref{e57}), consequently the width equation, effective potential, equilibrium width equation and the frequency of collective oscillations are given respectively by
\begin{equation}\label{e61}
\ddot{\xi_{r}}=\frac{1}{\xi_{r}^{3}}-\xi_{r}+\frac{P}{\xi_{r}^{4}}+\frac{K}{\xi_{r}^{7}}\,.
  \end{equation}

\begin{equation}\label{e62}
V(\xi_{r})=\frac{1}{2}(\xi_{r}^{2}+\frac{1}{\xi_{r}^{2}})+\frac{P}{3\xi_{r}^{3}}+\frac{K}{6\xi_{r}^{6}}\,.
\end{equation}

\begin{equation}\label{e63}
\frac{1}{\xi_{r0}^{3}}-\xi_{r0}+\frac{P}{\xi_{r0}^{4}}+\frac{K}{\xi_{r0}^{7}}=0\,.
                         \end{equation}

\begin{equation}\label{e64}\begin{split}\omega_{r}&=\sqrt{5-\frac{1}{\xi_{r0}^{4}}+\frac{3K}{2\xi_{r0}^{8}}}\\
&=\sqrt{8-\frac{4}{\xi_{r0}^{4}}-\frac{3P}{\xi_{r0}^{5}}}\,.\end{split}
\end{equation}
\noindent
where Eq.~(\ref{e63}) was used in Eq.~(\ref{e64}). Therefore, for the repulsive two-body interaction case, we have only one solution for Eq.~(\ref{e63}), then the condensate is always stable. As for the case of an attractive two-body interaction, we have a different situation: the equation can have no equilibrium position, or it could have up to three equilibrium solutions. The results are summarized in Fig.~6 as the variation of the equilibrium width in term of the interaction parameters obtained by solving numerically Eq.~(\ref{e63}).
\begin{figure}[t]
\begin{minipage}{.5\linewidth}
 \centerline{\psfig{file=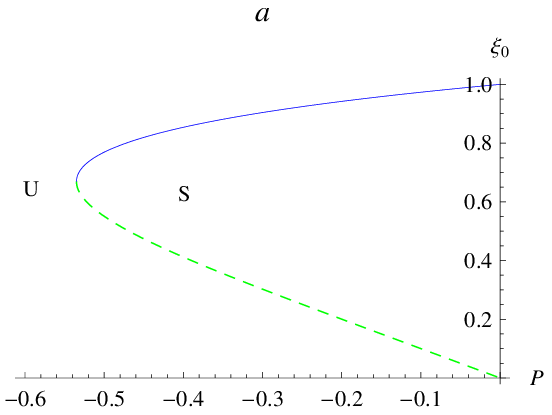,width=2.5in}}
\end{minipage}
\hspace{0.8cm}
\begin{minipage}{.4\linewidth}
 \centerline{\psfig{file=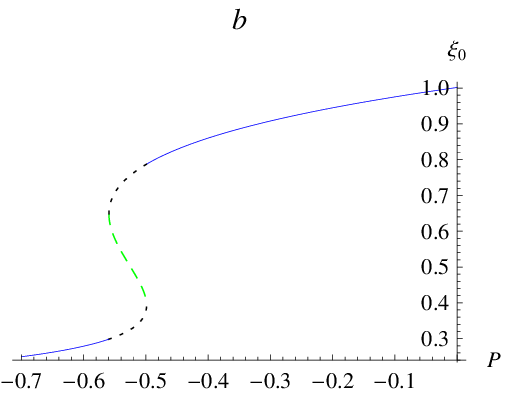,width=2.5in}}
\end{minipage}

\begin{center}\begin{minipage}{.4\linewidth}
 \centerline{\psfig{file=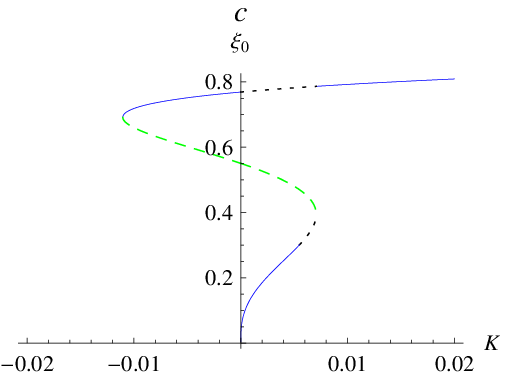,width=2.5in}}
\end{minipage}\end{center}

\vspace*{8pt}\caption{Condensate width $\xi_{r0} = \xi_{z0} = \xi_{0}$ for (a) $K = 0$, as a function of $P$; (b) $K = 0.007$, as a function of $P$; (c) $P = -0.5$, as a function of $K$. Solid blue lines represent the
stable solution, dotted black lines represent another stable solution, and dashed green lines represent unstable solution.}
\end{figure}
Fig.~6(a) shows solutions for $K = 0$ as a function of $P$. As we see, the solid blue curve represents the stable solution, whereas, the dashed green one depicts the unstable solution. In addition the critical point is identified as the join of the two curves, its value is $P_{c}=-0.535$ which coincides exactly with the value from Ref.~\cite{r7}. Between the two curves we have the stable area (S) such as if the initial width lies inside (S) the collapse is avoided, otherwise the collapse occurs (region U) even if the interaction strength is still finite.

Next, considering the combination of an attractive two-body interaction and a small repulsive three-body
interaction, we get different results of the stability. The system can either have one or
three solutions, as shown in Fig.~6(b). Despite being small, the addition of a positive three-body interaction $K$  leads to the existence of at least one stable solution for each value of $P$. Nevertheless, as we can see from Fig.~6(b), for large negative values of $P$  the curve representing $\xi_{0}(P)$  is flat, which means that the stability region can be considerably enhanced by the inclusion of a small positive value of $K$ compared to the case of pure two-body interaction. Furthermore, if the negative values of $P$ is large enough, the equilibrium width tends to zero, which practically represents a collapsed condensate.

In Fig.~6(c) we see that the stable solution for $P =-0.5$ exists only for a limited interval of negative values of $K$ which leads to the fact that both attractive three-body and two-body interactions cause the same effect of instability on the BEC.

Now let us explain the differences between the behavior of the BEC with logarithmic interaction and the BEC with two-body and three-body interactions. First of all, we begin with the most interesting cases of positive $\lambda$ and negative $P$ (with $K=0$ ), it is clear that both Eq.~(\ref{e18}) and Eq.~(\ref{e61}) express otherwise the dynamics of one-dimensional soliton confined in harmonic trap. It is well known that in addition to the attractive term caused by the trap, such soliton objects are formed by a compensation between a dispersive and non-linear effects. In the two equations the dispersive effect is repulsive and proportional to $\xi^{-3}$, but the non-linear effects are both attractive and proportional to $\xi^{-1}$ for Eq.~(\ref{e18}) and $\xi^{-4}$ for Eq.~(\ref{e61}). The remarkable difference between the two systems will be more apparent if $\xi\rightarrow 0$. In fact, at this limit and for a small $\lambda$ the dispersive term $\xi^{-3}$ dominates the non-linear one $ \lambda \xi^{-1}$ for the soliton with logarithmic nonlinearity (Eq.~(\ref{e18})). Therefore, as mentioned above the system remains stable and the collapse can take place only if $\lambda$ gets larger, i.e the non-linear term becomes larger. However, for Eq.~(\ref{e61}) the non-linear term $ P \xi^{-4}$ dominates the dispersive one $\xi^{-3}$, and the above compensation is no longer possible.  Consequently, the BEC with two-body interaction collapses at the center of the potential. Yet, the addition of a small repulsive three-body interaction (depending on $K \xi^{-7}$) to the pure two-body one tends to push the soliton away from the center and to extend the region of stability, but even the stability region of the BEC with an attractive two-body interaction can be considerably extended by the addition of a small repulsive three-body interaction, it never reaches that related to the BEC with logarithmic interaction, and this fact remains one of the most discriminating features between the two systems.
\section{Conclusion}

The Gaussian variational approach is used to determine the dynamics of an hypothetical BEC governed by the so-called logarithmic Schrodinger equation. Despite being non-linear, the evolution equation for the width has provided a great simplification of the original problem, and its analytical solution has been verified to be in a good agreement with numerical results. This equation has demonstrated that the condensate width evolves like a classical particle in anharmonic potential.

The stability of a trapped logarithmic condensate studied in the present work by considering two important quantities: the collective oscillations frequency and the mean-square radius. Based on the relevant requirements for instability which gives us the critical points for collapse and which states that the BEC becomes unstable if the oscillation frequency is zero or alternatively the mean-square radius of the condensate wave-function tends to zero in a finite time, it has been shown that our condensate in isotropic harmonic trap is always stable, and that the collapse is possible if the logarithmic strength is positive and sufficiently large. For the case of axially harmonic trap, it is proved that once the anisotropy gets larger, the equilibrium width tends to zero for either positive or negative values of logarithmic parameter. Consequently, beside the effect of logarithmic strength, there is also the effect of anisotropy which can promote collapse, in other words this phenomenon occurs also as the condensate takes a very thin pancake-shape.

Finally, the comparison between the BEC with logarithmic interaction and the BEC with two-body and three-body interactions has definitely demonstrated that the former is more stable than the latter.

\end{document}